\documentclass[11pt,a4paper]{article}
\pdfoutput=1

\usepackage[utf8]{inputenc}
\usepackage[margin=1in]{geometry}
\usepackage{authblk}
\usepackage[style=apa, uniquename=false]{biblatex}
\usepackage[american]{babel}
\DeclareLanguageMapping{american}{american-apa}
\bibliography{bibliography.bib}

\usepackage{amsmath,amsfonts,amssymb,bm,bbm,amsthm}
\usepackage{graphicx}
\usepackage{nicefrac}

\usepackage[colorlinks=true, allcolors=blue, bookmarks=false]{hyperref}

\providecommand{\keywords}[1]
{
  \small	
  \textbf{Keywords:} #1
}

\newcommand{\FAppel}[6]{\,F_1\left({#1}, {#2}, {#3}; {#4}; {#5}, {#6}\right)}
\newcommand{\RV}[1]{\textcolor{black}{#1}}

\date{\small{\textcolor{blue}{This manuscript has been accepted for publication in \textit{Statistics in Medicine}}}.}
\title{A Puzzle of Proportions: Two Popular Bayesian Tests Can Yield Dramatically Different Conclusions}

\author[1]{Fabian Dablander}
\author[1]{Karoline Huth}
\author[1]{Quentin F. Gronau}
\author[2]{Alexander Etz}
\author[1]{Eric-Jan Wagenmakers}
\affil[1]{Department of Psychological Methods, University of Amsterdam}
\affil[2]{University of California, Irvine}

\begin{document}
\maketitle

\begin{abstract}
\noindent Testing the equality of two proportions is a common procedure in science, especially in medicine and public health. In these domains it is crucial to be able to quantify evidence for the absence of a treatment effect. Bayesian hypothesis testing by means of the Bayes factor provides one avenue to do so, requiring the specification of prior distributions for parameters. The most popular analysis approach views the comparison of proportions from a contingency table perspective, assigning prior distributions directly to the two proportions. Another, less popular approach views the problem from a logistic regression perspective, assigning prior distributions to logit-transformed parameters. Reanalyzing 39 null results from the \textit{New England Journal of Medicine} with both approaches, we find that they can lead to markedly different conclusions, especially when the observed proportions are at the extremes (i.e., very low or very high). We explain these stark differences and provide recommendations for researchers interested in testing the equality of two proportions and users of Bayes factors more generally. \RV{The test that assigns prior distributions to logit-transformed parameters creates prior dependence between the two proportions and yields weaker evidence when the observations are at the extremes. When comparing two proportions, we argue that this test should become the new default.}
\end{abstract}

\keywords{Bayesian testing, Evidence of absence, Equality of proportions, Data reanalysis}

\maketitle

\section{Introduction}
Researchers frequently wish to test whether two populations differ. In medicine and public health, for example, the resulting statistical analysis frequently concerns testing whether or not two proportions differ. Examples include testing whether a vaccine decreases the number of infections compared to a control \parencite[e.g.,][]{polack2020safety}, whether sexual minorities are more prone to suicide compared to their heterosexual counterparts \parencite[e.g.,][]{ploderl2013suicide}, or whether tightly or less-tightly controlling hypertension leads to fewer miscarriages in pregnant women \parencite{magee2015less}.

In these applications it is crucial to be able to discriminate between evidence of absence and absence of evidence. For example, \textcite{magee2015less} conducted a trial to investigate the effect of a tight (target diastolic blood pressure, 85 mm Hg) or a less-tight (target diastolic blood pressure, 100 mm Hg) control of hypertension in pregnant woman on, among other outcomes, pregnancy loss. They found no significant difference between the two conditions, with 15 out of 493 women in the less-tight control condition and 13 out of 488 in the tight control condition having lost their child, yielding an estimated odds ratio of 1.14 (95\% CI: [0.53, 2.45]). How confident are we that there is indeed no difference between the two conditions rather than the data being inconclusive? Bayesian statistics provides a principled way of quantifying evidence via the Bayes factor \parencite{jeffreys1935some, ly2016harold, morey2016philosophy}, thus providing one avenue to discriminate between evidence of absence and absence of evidence \parencite[e.g.,][]{keysers2020using}.

The Bayes factor quantifies how well one hypothesis predicts the data compared to another. Using the test of equality between two proportions as an example, let $\mathcal{D} = (y_1, y_2, n_1, n_2)$ denote the combined data from the two groups. We have:
\begin{align*}
    Y_1 &\sim \text{Binomial}(n_1, \theta_1) \\
    Y_2 &\sim \text{Binomial}(n_2, \theta_2) \enspace ,
\end{align*}
where the sample sizes $(n_1, n_2)$ are assumed fixed and under $\mathcal{H}_0$ we have that $\theta \equiv \theta_1 = \theta_2$ while under $\mathcal{H}_1$ we have that $\theta_1 \neq \theta_2$. By quantifying relative predictive performance, the Bayes factor tells us how we should update our prior beliefs about $\mathcal{H}_0$ relative to $\mathcal{H}_1$ after observing the data \parencite{kass1995bayes}:
\begin{equation*}
    \underbrace{\frac{p(\mathcal{H}_0 \mid \mathcal{D})}{p(\mathcal{H}_1 \mid \mathcal{D})}}_{\text{Posterior odds}} = \underbrace{\frac{p(\mathcal{D} \mid \mathcal{H}_0)}{p(\mathcal{D} \mid \mathcal{H}_1)}}_{\text{Bayes factor}} \, \times \underbrace{\frac{p(\mathcal{H}_0)}{p(\mathcal{H}_1)}}_{\text{Prior odds}} \enspace .
\end{equation*}
\RV{A Bayes factor of, say, 15 means that the data are 15 times more likely under one hypothesis compared to the other. While there exist verbal guidelines that may aid in the interpretation of the Bayes factor \parencite[for example, Bayes factors in the range from 1$-$3 constitute \emph{weak} evidence, those in the range from 3$-$10 constitute \emph{moderate} evidence, and values larger than $10$ constitute \emph{strong} evidence;][]{jeffreys1939theory, wasserman2000bayesian, lee2013book}, the Bayes factor should be understood as a continuous measure of evidence \parencite{morey2016philosophy}}.

While the Bayes factor does not depend on the prior probability of hypotheses, it does depend crucially on the prior over parameters in the models instantiating $\mathcal{H}_0$ and $\mathcal{H}_1$, which becomes apparent when expanding:
\begin{equation*}
    \frac{p(\mathcal{D} \mid \mathcal{H}_0)}{p(\mathcal{D} \mid \mathcal{H}_1)} = \frac{\int_{\theta} p(\mathcal{D} \mid \theta, \mathcal{H}_0) \pi_0(\theta \mid \mathcal{H}_0) \, \, \mathrm{d} \theta}{\int_{\theta_1} \int_{\theta_2} p(\mathcal{D} \mid \theta_1, \theta_2, \mathcal{H}_1) \pi_1(\theta_1, \theta_2 \mid \mathcal{H}_1) \, \, \mathrm{d} \theta_1 \, \mathrm{d} \theta_2} \enspace ,
\end{equation*}
where $\pi_0$ and $\pi_1$ indicate the respective prior distributions.

There exist two main Bayes factor approaches for testing the equality of two proportions. The more popular one comes from the analysis of contingency tables, and assigns independent beta distributions directly to $(\theta_1, \theta_2)$ \parencite[e.g.,][]{gunel1974bayes, jamil2017default}. We call this approach the ``Independent Beta'' (IB) approach. The second approach is less widely used, and assigns a prior to the average log odds $\beta$ and the log odds ratio $\psi$ \parencite{kass1992approximate, gronau2019informed}. We call this approach the ``Logit Transformation'' (LT) approach. In this paper, we show that these two approaches can yield markedly different results. This is especially the case when the observed proportions are at the extremes (i.e., very low or very high), as is the case for a large number of applications including the three examples mentioned above. Consider the study by \textcite{magee2015less} again. The IB approach yields a Bayes factor of 12.30 in favour of $\mathcal{H}_0$, while the LT approach yields a mere 1.17. In other words, under the IB approach the data are about 12 times more likely under the hypothesis that tightly or less-tightly controlling hypertension have the same effect on miscarriages compared to a hypothesis assuming a difference. Under the LT approach, however, the data are about equally likely under both hypotheses, which constitutes equivocal evidence. The answer to the question ``is observing two equally small proportions strong or weak evidence for the null hypothesis?'' depends, therefore, crucially --- and non-trivially --- on the prior setup.

This paper is structured as follows. In Section \ref{sec:two-tests}, we outline these two ways of testing the equality of two proportions in more detail. In Section \ref{sec:paradox}, we highlight the occasionally stark differences of the two approaches by reanalyzing 39 statistical tests reported in the \textit{New England Journal of Medicine} and explain why these differences occur. In Section \ref{sec:discussion}, we end by reviewing the implications of the prior setup and what users of Bayes factors should be mindful of when testing the equality of two population parameters. \RV{We argue that the LT approach should become the default when testing the equality of two proportions because it (a) induces prior dependence between proportions which almost always are, in fact, dependent, and (b) yields a sensibly milder assessment of the evidence compared to the IB approach when the observations are at the extremes.}

\section{Two Ways of Testing the Equality of Two Proportions} \label{sec:two-tests}
In this section, we outline two ways of testing the equality of two proportions. In Section \ref{sec:independent-beta} we describe the Independent Beta approach, and in Section \ref{sec:logit-transformation} we describe the Logit Transformation approach.

\subsection{The Independent Beta (IB) Approach} \label{sec:independent-beta}
In order to nest the null hypothesis under the alternative hypothesis, we introduce the difference parameter $\eta = \theta_2 - \theta_1$ and the grand mean $\zeta = \frac{1}{2} \left(\theta_1 + \theta_2\right)$. Using this parameterization, we have that:
\begin{align*}
    \theta_1 &= \zeta - \frac{\eta}{2} \\
    \theta_2 &= \zeta + \frac{\eta}{2} \enspace .
\end{align*}
The hypotheses are then specified as:
\begin{align*}
    \mathcal{H}_0&: \eta = 0 \\
    \mathcal{H}_1&: \eta \neq 0 \enspace .
\end{align*}

\begin{figure}[h!]
   \centering
   \includegraphics[width = 1\textwidth]{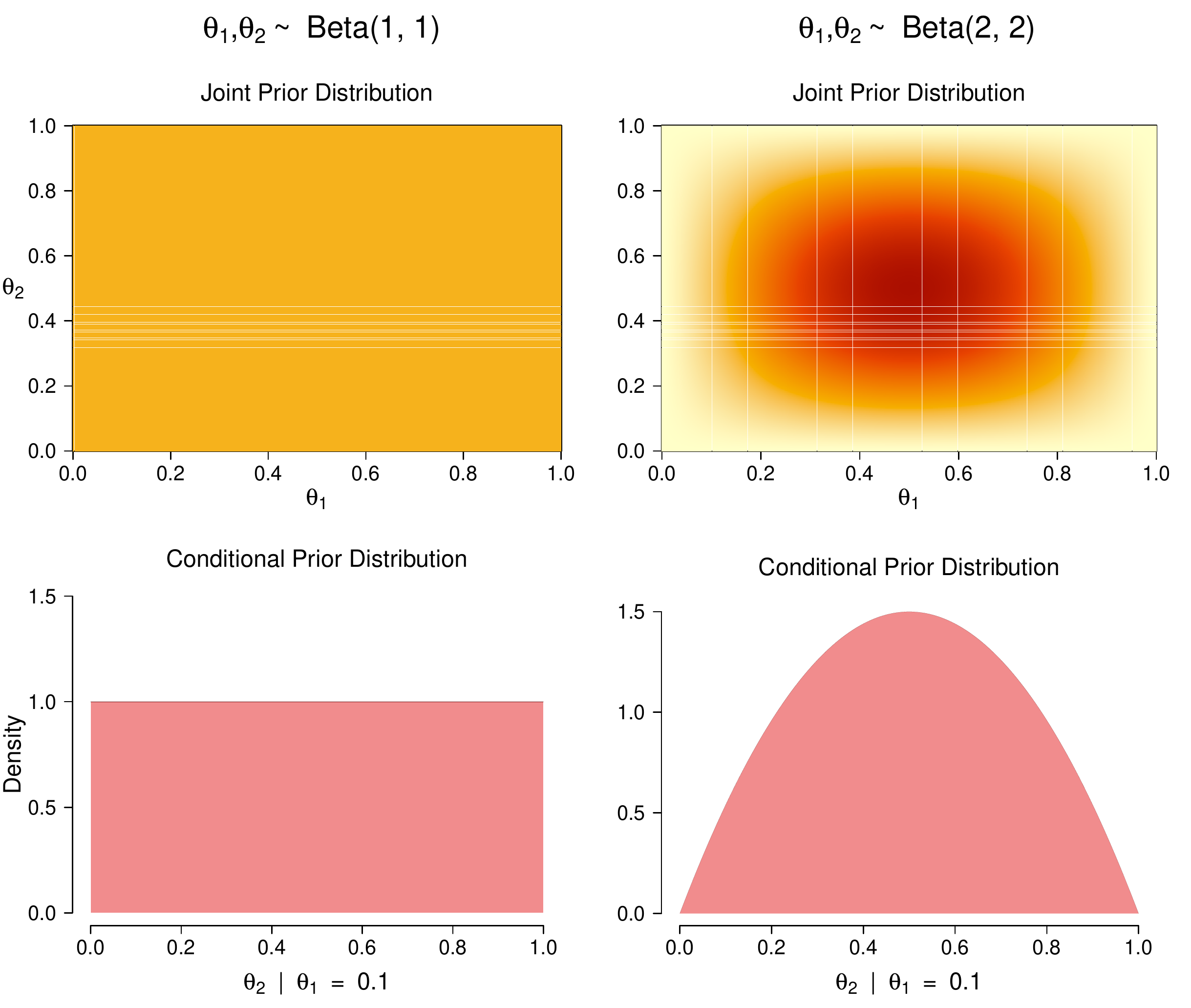}
   \caption{Top: Joint prior distribution assigned to $(\theta_1, \theta_2)$ under $a = 1$ (left) and $a = 2$ (right). Bottom: Conditional prior distribution of $\theta_2$ given that $\theta_1 = 0.10$.}
   \label{fig:independent-priors}
\end{figure}

Under this approach, we assign independent $\text{Beta}(a, a)$ priors to $\theta_1$ and $\theta_2$. Figure \ref{fig:independent-priors} visualizes the joint prior distribution (top) under $\mathcal{H}_1$ for $a = 1$ (left) and $a = 2$ (right). Increasing values of $a$ implies that the joint prior mass is more concentrated around $(\theta_1, \theta_2) = (\nicefrac{1}{2}, \nicefrac{1}{2})$. The bottom panels visualize the conditional prior distribution of $\theta_2$ given that we know that $\theta_1 = 0.10$. Knowing the value of $\theta_1$ does not change our prior about $\theta_2$, which follows from the assumption of prior independence.

The IB Bayes factor is available in analytic form \parencite[see e.g.,][]{dickey1970weighted, gunel1974bayes, jeffreys1939theory}. In the literature on contingency tables, our setup corresponds to an independent multinomial sampling scheme where the row (or column) sums of the contingency table (here $n_1$ and $n_2$) are fixed; for an extension to other sampling schemes and more than two groups, see \textcite{gunel1974bayes} and \textcite{jamil2017default}. \textcite{gunel1974bayes} suggest $a = 1$ as a default value. Note that values $a < 1$ lead to an undefined prior density for $\theta = 0$ and $\theta = 1$, which implies model selection inconsistency in case $\theta$ is indeed 1 or 0. Hence these values should be avoided on principle grounds. Because the beta prior is conjugate for the binomial likelihood, the posterior distributions of $\theta_1$ and $\theta_2$ are again (independent) beta distributions.

\subsection{The Logit Transformation (LT) Approach} \label{sec:logit-transformation}
The test proposed by \textcite{kass1992approximate} and implemented by \textcite{gronau2019informed} does not assign a prior directly to $(\theta_1, \theta_2)$, but applies a logit transformation and assigns priors to the transformed parameters ($\beta$, $\psi$). Specifically, we write:
\begin{align*}
    \text{log}\left(\frac{\theta_1}{1 - \theta_1}\right) &= \beta - \frac{\psi}{2} \\
    \text{log}\left(\frac{\theta_2}{1 - \theta_2}\right) &= \beta + \frac{\psi}{2} \enspace ,
\end{align*}
where $\beta$ is a grand mean and $\psi$ is the difference in log odds (i.e., the log odds ratio):
\begin{align*}
    \beta &= \frac{1}{2}\left(\text{log}\left(\frac{\theta_1}{1 - \theta_1}\right) + \text{log}\left(\frac{\theta_2}{1 - \theta_2}\right)\right) \\
    \psi &= \text{log}\left(\frac{\theta_2}{1 - \theta_2}\right) - \text{log}\left(\frac{\theta_1}{1 - \theta_1}\right) \enspace .
\end{align*}

While this is a more involved reparameterization than in the IB approach above, another way to formulate this setup is by writing:
\begin{align*}
    \theta_1 &= \frac{e^{\beta - \frac{\psi}{2}}}{1 + e^{\beta - \frac{\psi}{2}}} \\
    \theta_2 &= \frac{e^{\beta + \frac{\psi}{2}}}{1 + e^{\beta + \frac{\psi}{2}}} \enspace ,
\end{align*}
which readers familiar with logistic regression may recognize. Using this setup, we test the hypotheses:
\begin{align*}
    \mathcal{H}_0&: \psi = 0 \\
    \mathcal{H}_1&: \psi \neq 0 \enspace .
\end{align*}

\begin{figure}[ht]
   \centering
   \includegraphics[width = 1\textwidth]{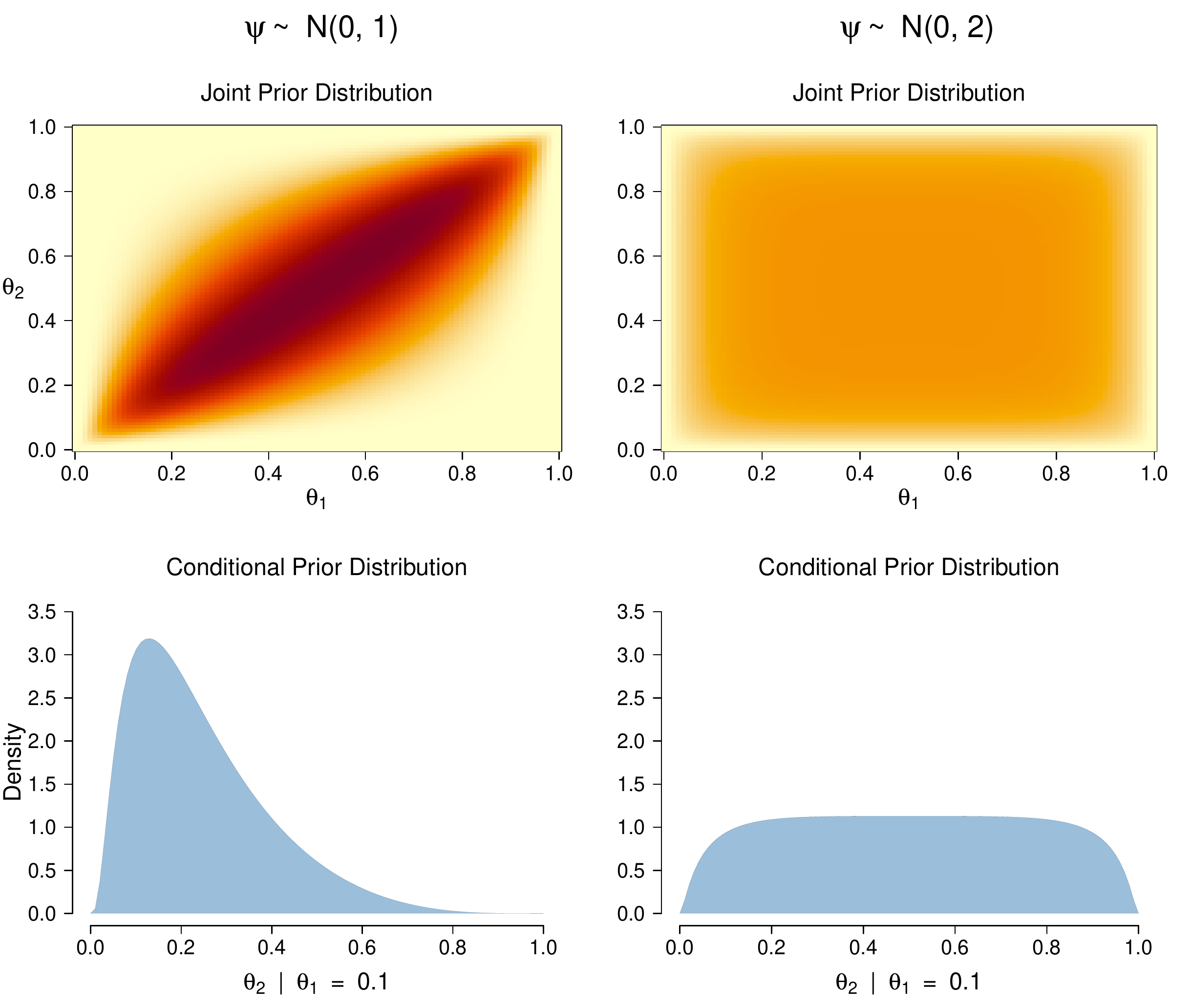}
   \caption{Top: Joint prior assigned to $(\theta_1, \theta_2)$ for $\sigma_{\psi} = 1$ (left) and $\sigma_{\psi} = 2$ (right). Bottom: Conditional prior distribution of $\theta_2$ given that $\theta_1 = 0.10$. In both cases we assume $\sigma_{\beta} = 1$.}
   \label{fig:dependent-priors}
\end{figure}

In contrast to above, we now assign priors to $\beta$ and $\psi$ rather than to $\theta_1$ and $\theta_2$ directly. In particular, under both hypotheses we assume $\beta \sim \mathcal{N}(0, \sigma_{\beta})$ with $\sigma_{\beta} = 1$. Under $\mathcal{H}_1$ we assume $\psi \sim \mathcal{N}(0, \sigma_{\psi})$.\footnote{While this prior specification assigns the log odds ratio $\psi$ a Gaussian distribution, the IB prior specification results in a non-standard distribution on $\psi$. For more details, see Appendix \ref{sec:induced-priors}.} The top left panel in Figure \ref{fig:dependent-priors} visualizes the implied joint prior distribution on $(\theta_1, \theta_2)$ under $\mathcal{H}_1$ for $\sigma_{\psi} = 1$. The prior mass is concentrated along the diagonal, which indicates that $\theta_1$ and $\theta_2$ are dependent. The bottom left panel illustrates this fact: if we know that $\theta_1 = 0.10$, then the prior on $\theta_2$ shifts toward this value. Setting $\sigma_{\psi} = 2$ removes the prior dependency, as the right column in Figure \ref{fig:dependent-priors} shows. For values of $\sigma_{\psi} > 2$, $\theta_1$ and $\theta_2$ become \textit{anti-correlated} and hence observing a small value of $\theta_1$ results in a prior that puts more mass on large values for $\theta_2$. Such an inverse relation is undesirable in almost all empirical applications, and so values $\sigma_{\psi} > 2$ are therefore to be avoided. \textcite{gronau2019informed} developed software to compute the Bayes factor using this prior specification, first proposed by \textcite{kass1992approximate}, suggesting $\sigma_{\psi} = 1$ as a default value.

\subsection{Comparison of Priors} \label{sec:prior-comparison}
A direct comparison of the two prior specifications may be helpful to get further intuition for their differences. While the IB approach does not assign a prior distribution to $(\beta, \psi)$ or $(\eta, \zeta)$ explicitly, assigning a prior to $\theta_1$ and $\theta_2$ induces a prior distribution on these quantities. Conversely, the LT approach assigns a prior to $\psi$ and $\beta$ and this induces a prior on $(\theta_1,\theta_2)$ and $(\eta, \zeta)$. The induced prior distributions under both approaches are non-standard (see Appendix \ref{sec:induced-priors}), but their densities can be calculated numerically.

\begin{figure}[ht]
   \centering
   \includegraphics[width = 1\textwidth]{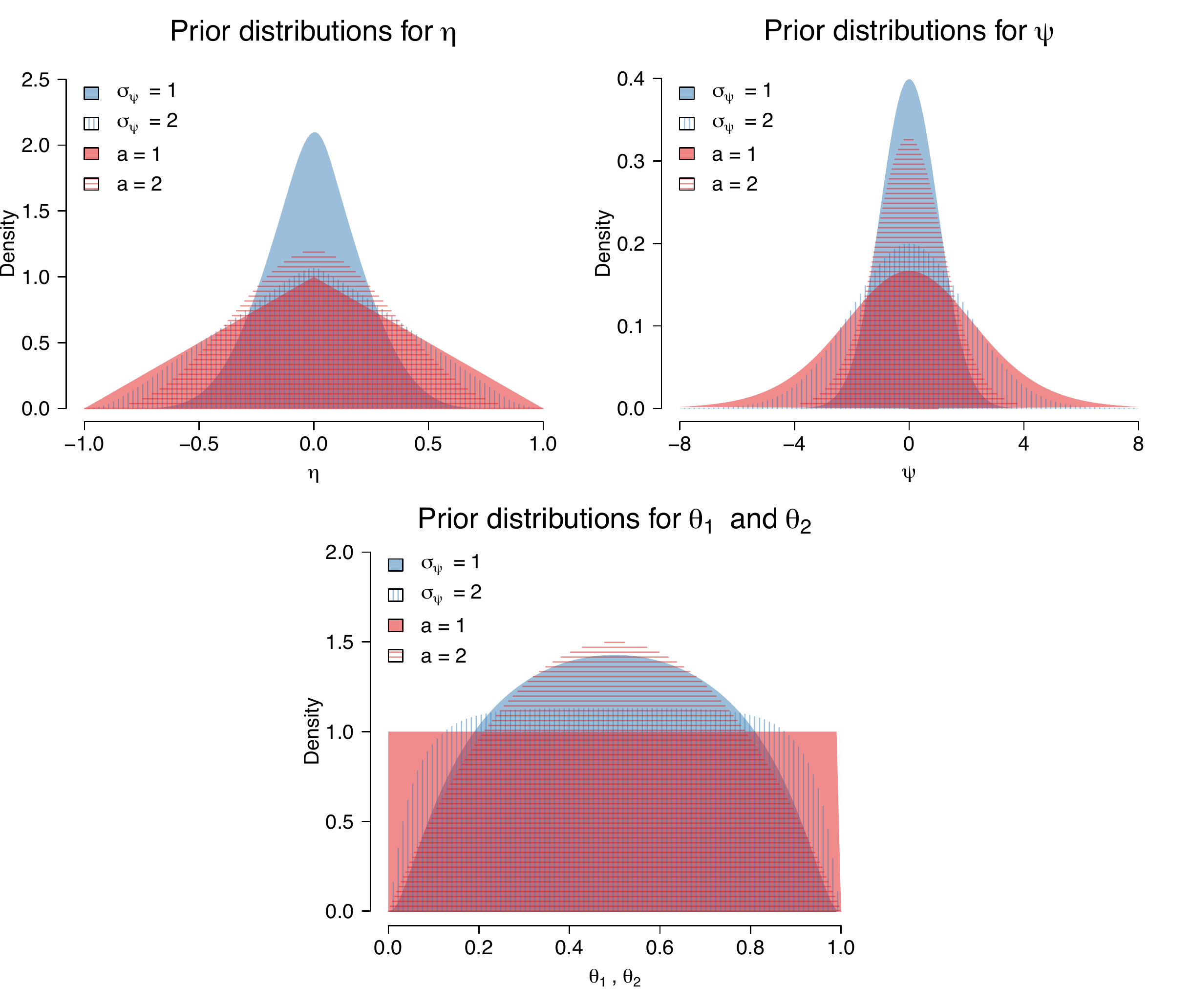}
   \caption{Top: Prior distributions assigned to $\eta$ (left) and $\psi$ (right) under the LT (blue, vertical lines) and IB approach (red, horizontal lines), for two different parameter settings, respectively. Bottom: Marginal prior distribution of $\theta_1$ and $\theta_2$ under the two approaches. \RV{The density that is filled and has the highest peak corresponds to $\sigma_{\psi} = 1$.}}
   \label{fig:prior-comparison}
\end{figure}

The top left panel in Figure \ref{fig:prior-comparison} shows the prior distribution assigned to $\eta$ by the LT approach for $\sigma_{\psi} = 1$ (shaded blue) and $\sigma_{\psi} = 2$ (striped blue) and the IB approach for $a = 1$ (shaded red) and $a = 2$ (striped red) under $\mathcal{H}_1$. Similarly, the top right panel shows the prior distribution assigned to $\psi$ for the two approaches and prior parameter values. The (default) IB approach assigns comparatively more mass to large values of $\eta$ and $\psi$, which in practice means that it expects larger differences between the sample proportions. The bottom panel shows the marginal priors for $\theta_1$ and $\theta_2$, where we find that the LT approach assigns comparatively less mass to extreme values. The LT approach cannot result in a uniform distribution on the proportions under $\mathcal{H}_0$ because of the Gaussian prior on $\beta$. If it instead would assign a (standard) logistic prior to $\beta$ (which has fatter tails), the prior on the proportions would be uniform; see a related discussion in Appendix \ref{sec:model-averaging}. In the next section, we discuss a somewhat surprising difference between these two tests.

\section{Practical Implications of the Prior Setup} \label{sec:paradox}
To see the implications of the two different prior specifications in practice, in Section \ref{sec:reanalysis} we reanalyze 39 null results published in the \textit{New England Journal of Medicine}, previously analyzed by \textcite{hoekstra2018bayesian} using the IB approach. We then explain why these difference occur in Section \ref{ref:explanation}. The data and code to reproduce all analyses and figures are available from \url{https://github.com/fdabl/Proportion-Puzzle}.

\subsection{Reanalysis of \textit{New England Journal of Medicine} Studies} \label{sec:reanalysis}
\begin{figure}[ht]
   \centering
   \includegraphics[width = 1\textwidth]{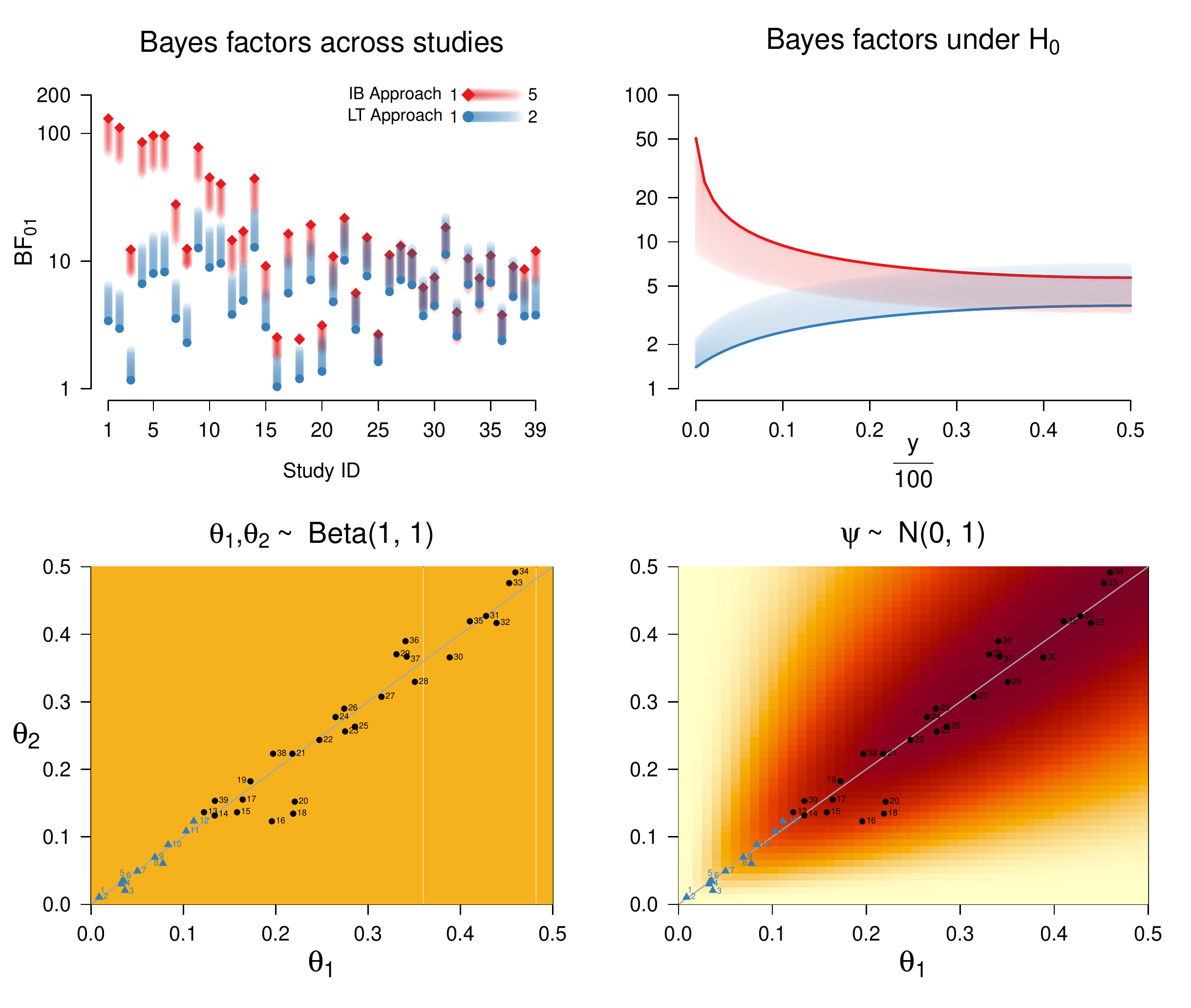}
   \caption{Top: Bayes factors using the IB (rectangles) and LT approach (circles) in favour of $\mathcal{H}_0$ across studies reported in \textcite{hoekstra2018bayesian} (left) or for simulated equal proportions with $n = 100$ (right) for values $a \in [1, 5]$ and $\sigma_{\psi} \in [1, 2]$ with $\sigma_{\beta} = 1$. Bottom: Joint prior distribution of $(\theta_1, \theta_2)$ under the IB approach with $a = 1$ (left) and under the LT approach with $\sigma_{\psi} = 1$ and $\sigma_{\beta} = 1$ (right). Black dots and blue rectangles indicate the maximum likelihood estimates of the proportions in the studies analyzed by \textcite{hoekstra2018bayesian}.}
   \label{fig:reanalysis-all}
\end{figure}

\textcite{hoekstra2018bayesian} considered all 207 articles published in the \textit{New England Journal of Medicine} in 2015. The abstract of 45 of these articles contained a claim about the absence or non-significance of an effect for a primary outcome measure, and 37 of those allowed for a comparison of proportions, reporting 43 null results in total. We focus on those results that can be reanalyzed using a test between two proportions, which results in a total of 39 tests from 32 articles. The top left panel in Figure \ref{fig:reanalysis-all} contrasts Bayes factors in favour of $\mathcal{H}_0$ computed using the IB approach (rectangles) across $a \in [1, 5]$ with Bayes factors computed using the LT approach (circles) across $\sigma_{\psi} \in [1, 2]$. In virtually all cases and across specifications, the Bayes factor in favour of $\mathcal{H}_0$ is higher under the IB approach, and this difference is frequently substantial.\footnote{In contrast, the conclusions one would draw based on posterior distributions are very similar, see Appendix \ref{sec:posterior-inference}.} As the parameter $a$ is increased under the IB approach, the expected difference between the two groups is smaller (see top left panel in Figure \ref{fig:prior-comparison}). Therefore, the predictions under $\mathcal{H}_1$ become more similar to the predictions under $\mathcal{H}_0$, and the Bayes factor decreases. Conversely, as $\sigma_{\psi}$ is increased under the LT approach, the expected difference between the two group increases, and the Bayes factor in favour of $\mathcal{H}_0$ increases.

The top left panel in Figure \ref{fig:reanalysis-all} shows that, for some studies, the IB and LT Bayes factors cannot be brought into the vicinity of each other by changing the prior parameters in the way specified above, while for other studies the Bayes factors do overlap substantially. The studies without overlap are those indexed as 1-12. They are shown as blue rectangles in the bottom panels in Figure \ref{fig:reanalysis-all}, which shows the joint prior density for the IB (left) and the LT approach (right) with the symbols indicating the maximum likelihood estimates for the individual studies. Note that the larger the proportions in the bottom panels, the more likely are the two Bayes factors to overlap in the top left panel.

While \textcite{hoekstra2018bayesian} are reassured by the fact that their Bayesian reanalysis (using the IB approach) yields strong evidence in favour of $\mathcal{H}_0$ on average (given that all studies reported a non-significant $p$-value), using the LT approach yields a more uncertain picture. While the median Bayes factor across the studies under the IB approach is 12.30, the median Bayes factor under the LT approach is only 4.79. This difference is driven by the extremes, where the two approaches suggest substantially different conclusions. We have already seen one example in the introduction concerning hypertension and pregnancy loss (analysis identifier 3). Another example is given by \textcite{joura20159}, which compared the efficacy of the 9vHPV against the qHPV vaccine for preventing, among others, cervical, vulvar, and varginal disease and persistent human papillomavirus related infections in women. Comparing different outcome measures in women who were HPV-negative and HPV-positive at baseline, the IB approach yields Bayes factors consistently around 100 while the LT approach never shows Bayes factors larger than 10 (see analysis identifiers 1, 2, 5, and 6). The IB approach thus strongly suggests that the two vaccines have a similar efficacy, while the LT approach suggests that more data is needed to reach a firm conclusion.

The initially diverging and then converging pattern of the Bayes factors can be seen neatly in the top right panel in Figure \ref{fig:reanalysis-all}, which shows the Bayes factors under the null hypothesis for increasing values of $\nicefrac{y}{100} \equiv \nicefrac{y_1}{100} = \nicefrac{y_2}{100}$. Note that the pattern is symmetric, so that for $\nicefrac{y}{100} \in [\nicefrac{1}{2}, 1]$ the same holds, just mirrored. At the extremes $\nicefrac{y}{n} \in \{0, 1\}$, the IB approach yields a Bayes factor in favour of $\mathcal{H}_0$ of 50.75, while the LT Bayes factor gives a mere 1.40. The Bayes factor \textit{decreases} as $\nicefrac{y}{n} \rightarrow \nicefrac{1}{2}$ under the IB approach, but \textit{increases} under the LT approach. Consequently, the difference between the Bayes factors becomes less pronounced as we move to more central proportions, which matches the observation in the empirical analysis shown in the top left panel in Figure \ref{fig:reanalysis-all}. For $\nicefrac{y}{n} = \nicefrac{1}{2}$, the IB Bayes factor yields 5.70 and the LT Bayes factor yields 3.67.

In summary, we have found that the two approaches give a very different answer to the question ``is observing two equally small proportions strong or weak evidence in favour of the null hypothesis?''. The IB Bayes factor suggests that the evidence is often orders of magnitude larger than the LT Bayes factor.

\subsection{Explaining the Difference} \label{ref:explanation}
Why does the IB Bayes factor yield so much stronger evidence than the LT Bayes factor for $\mathcal{H}_0$ when the proportions are small and of roughly equal size? And why does the IB Bayes factor \textit{decrease} as the proportions get closer to $\nicefrac{1}{2}$ while the LT Bayes factor \textit{increases}? To answer these questions, we need to zoom in on the differences in the respective prior specifications. The two approaches differ in two key ways. First, while the IB approach assigns \textit{independent} priors to $\theta_1$ and $\theta_2$, the LT approach assigns \textit{dependent} priors to them. Second, the LT approach employs a logit transformation while the IB approach does not. Appendix \ref{sec:dependent-CT} shows that it is not the prior dependence that underlies the difference between the two approaches. Instead, as we will see below, it is the logit transformation.

\begin{figure}[h!]
   \centering
   \includegraphics[width = 1\textwidth]{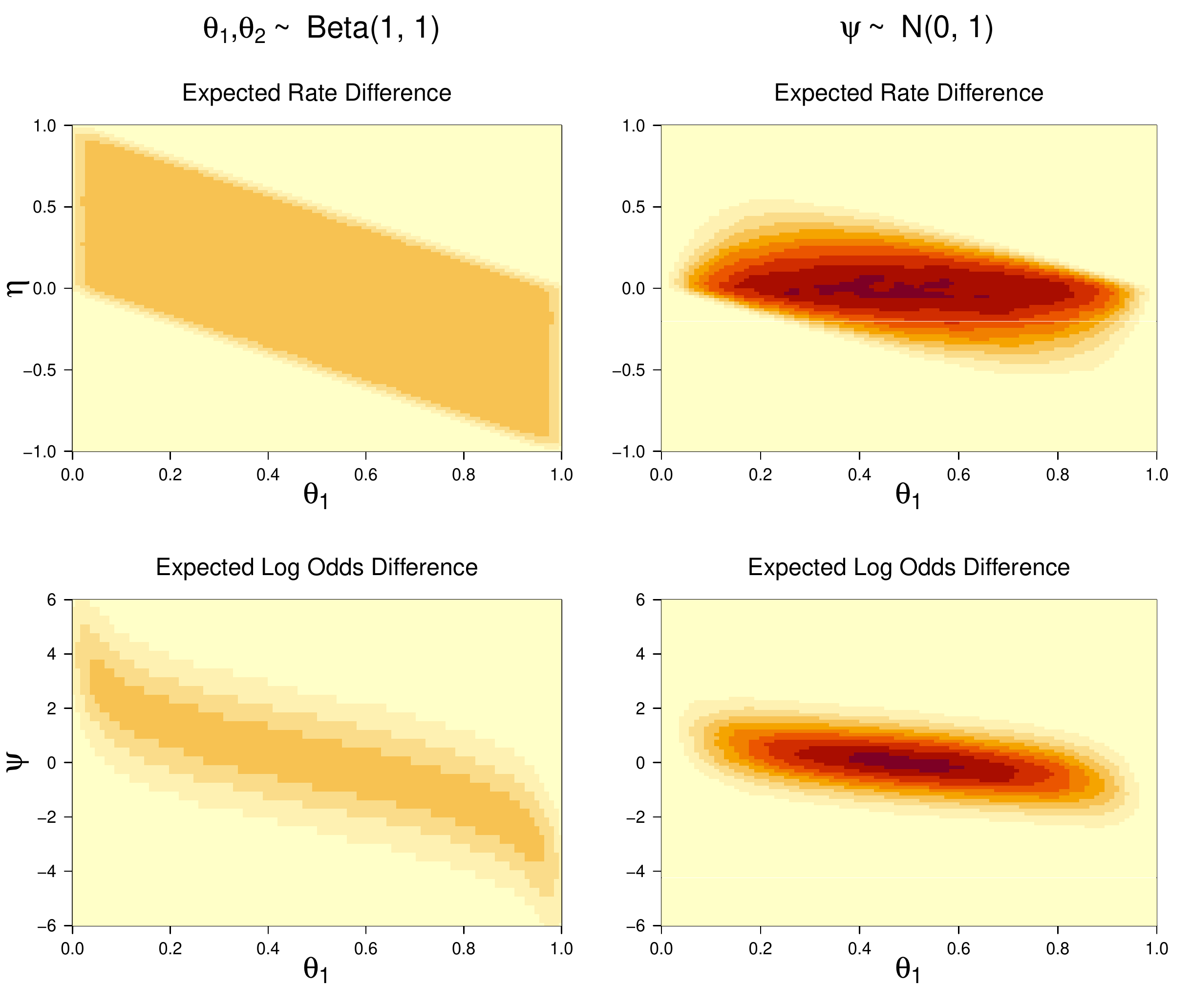}
   \caption{Top: Joint prior distribution for $(\theta_1, \eta)$ under the IB approach with $a = 1$ (left) and under the LT approach with $\sigma_{\psi} = 1$ and $\sigma_{\beta} = 1$ (right). Bottom: Joint prior distribution for $(\theta_1, \psi)$ under the two approaches.}
   \label{fig:updating-comparison}
\end{figure}

The fact that the IB Bayes factor is larger than the LT Bayes factor even when the latter approach expects a larger difference in the proportions (e.g., a larger $\eta$ or $\psi$, compare $a = 2$ with $\sigma_{\psi} = 2$ in Figure \ref{fig:prior-comparison}) means that focusing only on the difference parameter is not sufficient to explain the difference between the Bayes factors. Instead, we turn to a sequential predictive perspective. From such a perspective, we may first use the prior to predict the data from group one, update the prior to a posterior, and then predict the data from group two. One can rewrite the marginal likelihoods to make this sequential perspective apparent, see Appendix \ref{sec:prequential}. This perspective shows that there is a crucial difference in the predictions that the CT and LT approach make under $\mathcal{H}_1$. Under both approaches there is a common parameter $\theta$ under $\mathcal{H}_0$, and the prior assigned to $\theta$ gets updated by data from the first group, after which predictions about the second group are made. Under the alternative hypothesis, however, the IB approach implies that observing data from group one does not update our beliefs about likely values of $\theta_2$, and so data from group one cannot inform the subsequent prediction about group two. Under the LT approach, such information sharing does take place. This difference is shown in the top panels in Figure \ref{fig:updating-comparison}, which visualizes the joint prior distribution for $\theta_1$ and $\eta \equiv \theta_2 - \theta_1$ under the IB (left) and LT (right) approaches. We see that under the IB approach, learning about $\theta_1$ does not influence our predictions about likely differences between the two groups. In contrast, under the LT approach we find that when we learn that $\theta_1$ is either small or large, we expect small differences between the proportions compared to when we learn that $\theta_1$ is about half. This explains why the LT Bayes factor in favour of $\mathcal{H}_0$ \textit{increases} as we move from extreme values of $\nicefrac{y}{n}$ towards values around $\nicefrac{1}{2}$, as shown in the top right panel in Figure \ref{fig:reanalysis-all}: at the extremes, $\mathcal{H}_1$ expects smaller differences between $\theta_2$ and $\theta_1$, and hence it more closely resembles $\mathcal{H}_0$, leading to a more equivocal Bayes factor. Moving towards more central values, $\mathcal{H}_1$ expects larger differences and hence $\mathcal{H}_0$ outpredicts it by a larger margin.

While prior dependence is necessary to change one's beliefs about likely differences between the proportions upon learning $\theta_1$, the fact that the expected difference \textit{decreases} in the LT approach is due to the logit transformation. When the two rates $(\theta_1, \theta_2)$ are small, say $\theta_1 = 0.05$ and $\theta_2 = 0.10$, the log odds difference $\psi$ is large, in this case $\psi = 0.75$. The prior on $\psi$ in the LT approach renders such large log odds differences to be unlikely, which means that a smaller rate difference is expected. Conversely, if the rates are somewhere in the center, say $\theta_1 = 0.50$ and $\theta_2 = 0.55$, then the log odds difference $\psi$ is small, in this case $\psi = 0.20$. The prior on $\psi$ in the LT approach views such small log odds differences as likely; in fact, much larger ones are possible. Therefore, the LT approach expects larger rate differences when the rates are in the center. On the log odds scale, the expected difference does not change markedly as $\theta_1$ varies, as the bottom right panel in Figure \ref{fig:updating-comparison} shows.\footnote{The fact that there is barely any prior mass at extreme values of $\theta_1$ under the IB approach is due to the Gaussian prior on $\beta$, as discussed in Section \ref{sec:prior-comparison}.}

In contrast to the LT approach, the IB approach does not automatically reduce the expectations about likely differences as we learn that $\theta_1$ takes on extreme values. Taking slices at particular values for $\theta$ in the top left panel in Figure \ref{fig:updating-comparison} would show uniform distributions ranging from $-\theta_1$ to $1 - \theta_1$. Instead of reducing the size of an expected difference at the extremes, the IB approach actually \textit{amplifies} it. This occurs because of the boundedness of the rate parameters. In particular, if $\theta_1 = 0$, then the only values for which, say, $|\eta| < 0.10$ are $\theta_2 \in [0, 0.10]$. Differences in absolute magnitude between the two rates of up to $0.10$ if $\theta_1 = 0$ are therefore assigned a probability of $0.10$. If, however, $\theta_1 = 0.50$, the values of $\theta_2$ for which $|\eta| < 0.10$ are $\theta_2 \in [0.40, 0.60]$. This means that differences in absolute magnitude between the two rates of up to $0.10$ if $\theta_1 = 0.50$ are assigned a probability of $0.20$, or twice as much as in the case that $\theta_1 = 0$. Another way to see this is to note that the expected difference in the case that $\theta_1 = 0$ is $\eta = 0.50$, while the expected difference in the case that $\theta_1 = 0.50$ is $\eta = 0$; more generally, the expected difference increases linearly as we move to the extremes, as can be seen in the top left panel in Figure \ref{fig:updating-comparison}. The bottom left panel in Figure \ref{fig:updating-comparison} shows an associated and pronounced increase in the expected log odds difference as $\theta_1$ moves towards extreme values.

\section{Discussion} \label{sec:discussion}
Being able to discriminate between evidence of absence and absence of evidence is key in medicine and public health. The Bayes factor, which pits the predictive performance of two hypotheses against each other, is a method for doing so. Reanalyzing 39 null results published in the \textit{New England Journal of Medicine}, we have found that the strength of evidence depends crucially on the prior specification. Comparing an approach that assigns independent beta distributions directly to the rate parameters (IB) to one that employs a logit transformation (LT), we have found that the former approach suggests evidence that is orders of magnitudes larger than the latter approach when observing small proportions of roughly equal size. Consider again the effect of tightly or less-tightly controlling hypertension in pregnant women \parencite{magee2015less}. The IB approach suggests a Bayes factor about ten times larger in favour of the null hypothesis than the LT approach. Similarly, the Bayes factor in favour of the hypothesis that two vaccines against the human papillomavirus are equally effective \parencite{joura20159} is up to 38 times larger under the IB approach, depending on the outcome measure. While the original statistical analysis by the authors is more involved (e.g., adjusting for various covariates), these differences are striking. As we have seen, they occur because of the boundedness of the parameter space, which leads to the expectations of large differences at the extremes under the IB approach. In contrast, the LT approach expects smaller differences at the extremes.

Although the two Bayes factors frequently differed markedly in size, they always \textit{both} provided evidence in favour of $\mathcal{H}_0$ in the cases we studied in this paper. This need not be the case, however. For example, a large-scale investigation into the effect of aspirin found that 26 out of $n_1 = 11,034$ who received a placebo suffered a fatal heart attack, but only 10 out of $n_2 = 11,037$ who received aspirin did \parencite{steering1989final}. The IB Bayes factor yields strong evidence in favour of $\mathcal{H}_0$ ($\text{BF}_{01} = 19.78$) while the LT Bayes factor finds moderate evidence in favour of $\mathcal{H}_1$ ($\text{BF}_{10} = 5.36$). This is again owed to the fact that the IB approach expects larger differences than the LR approach; a small effect size is more likely under $\mathcal{H}_0$ than under $\mathcal{H}_1$ in the IB approach simply because under $\mathcal{H}_1$ unrealistically large effect sizes are assumed to be plausible.

Our results have a number of implications. First, researchers should think carefully about their prior setup. This not only implies thinking about the parameter values for the priors, but also the nature of the priors and what they imply for the data. As we have seen, one may be fooled by assessing how sensitive the Bayes factor is by varying only prior parameters conditional on the model. For all studies we re-analyzed that had proportions at the extremes, varying the prior parameters of the IB approach still resulted in substantial evidence in favour of the null hypothesis. However, changing the prior setup to the LT approach, we found markedly reduced evidence. If there are sensible alternative parameterizations, it may be prudent to explore how sensitive the results are to these. In particular, our exposition suggests that assessing the predictions that follow from the model can help get a better intuition of its assumptions. In our case, this made clear that the IB approach expects larger rate differences at the boundaries of the parameter space compared to the center, while the LT approach expects the reverse. We were struck by the initially puzzling contrast between these two tests, given that the task --- comparing two proportions --- seemed so simple. This suggests that even more caution needs to be applied when using Bayes factors for comparing models that are invariably more complex. In the same spirit of sensitivity analysis, it is prudent to compare inferences based on the Bayes factor with inferences based on the posterior distributions of the parameters. While testing precise hypotheses based on credible intervals is arguably inappropriate since it makes inference conditional on $\mathcal{H}_1$, ignoring $\mathcal{H}_0$ \parencite[e.g.,][]{berger2006bayes, wagenmakers2020principle}, alarm bells should go off when they show stark differences to the Bayes factor results, as in our case (compare Figure \ref{fig:reanalysis-all} and \ref{fig:ci-reanalysis}). \RV{Researchers who rely only on the posterior distribution for inference or for trial design are unaffected by our results.}

The second implication of our work is that the LT approach appears better suited for testing the equality of two proportions than the currently more popular IB approach. First, assuming prior dependence strikes us as a more sensible approach especially for medicine and public health. This is because it is generally unlikely that gaining information about one group does not influence our beliefs about the other group. For example, if a particular treatment for cancer yields a $30\%$ success rate, surely this should inform our expectations for the success rate of a comparison treatment. Incorporating this dependency results in more cautious predictions, which reduces the eagerness with which the standard IB approach suggests strong evidence in favour of $\mathcal{H}_0$ (see also Appendix \ref{sec:dependent-CT}). While we have focused on testing the sharp null hypothesis $\theta_1 = \theta_2$, the difference between dependent and independent priors is important also when testing other hypotheses. \textcite{howard19982} discusses the difference between a dependent and independent prior in the context of comparing $\theta_1 < \theta_2$ against $\theta_1 > \theta_2$, and similarly recommends dependent priors, as independent priors lead to a test that is not ``sufficiently cautious''.

Most importantly, however, is the fact that the LT approach avoids the issues associated with a bounded parameter space. In particular, the IB approach is not sensitive to the fact that the implications of a rate difference $\theta_2 - \theta_1$ depends on the actual values of the rates. For example, $\theta_2 = 0.10$ and $\theta_1 = 0.05$ and $\theta_2 = 0.55$ and $\theta_1 = 0.50$ imply the same rate difference, but while the former represents a 100\% increase, the latter represents only a 10\% increase. Yet the IB approach assigns more prior mass to large rate differences at the extremes of the parameter space compared to its center. In contrast, the reverse holds for the LT approach because of the logit transformation. This strikes us as more sensible. With the recent development of accessible software that makes the LT approach easily available \parencite{gronau2019informed, hoffmann2021bayesian}, we hope that the use of the LT approach for testing the equality of two proportions becomes more widespread.

To sum up, we have seen an initially puzzling divergence in evidence when reanalyzing 39 null results from the \textit{New England Journal of New Medicine} with two different Bayes factor tests. We have explained this divergence and sided with the approach that employs a logit transformation and assumes prior dependence between the rates. We suggest that this approach, rather than the one motivated from a contingency table perspective, should be routinely applied when testing the equality of two proportions. Our journey should also act as a cautionary tale for users of Bayes factors, stressing the importance of assessing the predictions of one's models, conducting thorough sensitivity analyses, and never relying on just a single quantity. We have demonstrated the usefulness of this in the simple case of testing the equality of two proportions. Most applications will arguably be more complex, however, increasing the potential for puzzles and stressing the need for a holistic evidence assessment.

\paragraph{Author contributions.} EJW proposed the study. KH, QFG, AE, and EJW worked on the problem initially. After an interim period, FD and EJW resumed the work and solved the puzzle. FD and KH analyzed the data. FD wrote the manuscript. All authors read, provided comments, and approved the submitted version of the paper. They also declare that there were no conflicts of interest.

\paragraph{Acknowledgements.} We thank Rink Hoekstra, Rei Monden, and Don van Ravenzwaaij for sharing the \textit{New England Journal of Medicine} data with us.


\printbibliography

@article{gronau2019informed,
  title={{Informed Bayesian inference for the A/B test}},
  author={Gronau, Quentin F and Raj, A and Wagenmakers, Eric-Jan},
  journal={Journal of Statistical Software},
  year={in press}
}

@article{ojo2003remark,
  title={A remark on the convolution of the generalized logistic random variables},
  author={Ojo, Matthew Oladejo},
  journal={ASSET serves A},
  volume={1},
  number={2},
  year={2003}
}

@article{kass1992approximate,
  title={{Approximate Bayes factors and orthogonal parameters, with application to testing equality of two binomial proportions}},
  author={Kass, Robert E and Vaidyanathan, Suresh K},
  journal={Journal of the Royal Statistical Society: Series B (Methodological)},
  volume={54},
  number={1},
  pages={129--144},
  year={1992},
  publisher={Wiley Online Library}
}

@BOOK{lee2013book,
  AUTHOR =       {Lee, M. D. and Wagenmakers, E.--J.},
  TITLE =        {Bayesian Cognitive Modeling: {A} Practical Course},
  PUBLISHER =    {Cambridge University Press},
  YEAR =         {2013}
}

@article{wasserman2000bayesian,
  title={Bayesian model selection and model averaging},
  author={Wasserman, Larry},
  journal={Journal of Mathematical Psychology},
  volume={44},
  number={1},
  pages={92--107},
  year={2000},
  publisher={Elsevier}
}

@article{hoeting1999bayesian,
  title={{Bayesian model averaging: A tutorial}},
  author={Hoeting, Jennifer A and Madigan, David and Raftery, Adrian E and Volinsky, Chris T},
  journal={Statistical Science},
  pages={382--401},
  year={1999},
  publisher={JSTOR}
}

@article{hinne2020conceptual,
  title={{A conceptual introduction to Bayesian model averaging}},
  author={Hinne, Max and Gronau, Quentin F and {van den Bergh}, Don and Wagenmakers, Eric-Jan},
  journal={Advances in Methods and Practices in Psychological Science},
  volume={3},
  number={2},
  pages={200--215},
  year={2020},
  publisher={SAGE Publications Sage CA: Los Angeles, CA}
}

@article{jamil2017default,
  title={{Default “Gunel and Dickey” Bayes factors for contingency tables}},
  author={Jamil, Tahira and Ly, Alexander and Morey, Richard D and Love, Jonathon and Marsman, Maarten and Wagenmakers, Eric-Jan},
  journal={Behavior Research Methods},
  volume={49},
  number={2},
  pages={638--652},
  year={2017},
  publisher={Springer}
}

@article{dickey1970weighted,
  title={{The weighted likelihood ratio, sharp hypotheses about chances, the order of a Markov chain}},
  author={Dickey, James M and Lientz, BP},
  journal={The Annals of Mathematical Statistics},
  pages={214--226},
  year={1970},
  publisher={JSTOR}
}

@article{hoekstra2018bayesian,
  title={{Bayesian reanalysis of null results reported in medicine: Strong yet variable evidence for the absence of treatment effects}},
  author={Hoekstra, Rink and Monden, Rei and {van Ravenzwaaij}, Don and Wagenmakers, Eric-Jan},
  journal={PloS One},
  volume={13},
  number={4},
  pages={e0195474},
  year={2018},
  publisher={Public Library of Science San Francisco, CA USA}
}

@article{howard19982,
  title={{The 2 $\times$ 2 table: A discussion from a Bayesian viewpoint}},
  author={Howard, JV},
  journal={Statistical Science},
  pages={351--367},
  year={1998},
  publisher={JSTOR}
}

@article{pham1993bayesian,
  title={Bayesian analysis of the difference of two proportions},
  author={Pham-Gia, Thu and Turkkan, Noyan and Eng, P},
  journal={Communications in Statistics-Theory and Methods},
  volume={22},
  number={6},
  pages={1755--1771},
  year={1993},
  publisher={Taylor \& Francis}
}

@article{keysers2020using,
  title={{Using Bayes factor hypothesis testing in neuroscience to establish evidence of absence}},
  author={Keysers, Christian and Gazzola, Valeria and Wagenmakers, Eric-Jan},
  journal={Nature Neuroscience},
  volume={23},
  number={7},
  pages={788--799},
  year={2020},
  publisher={Nature Publishing Group}
}

@article{hoffmann2021bayesian,
  title={{Bayesian Inference for the A/B Test: Example Applications with R and JASP}},
  author={Hoffmann, Tabea and Wagenmakers, Eric-Jan},
  year={2021},
  doi={10.31234/osf.io/z64th},
  publisher={PsyArXiv}
}

@INCOLLECTION{berger2006bayes,
  author =     {Berger, J. O.},
  year =       {2006},
  editor =     {Kotz, S. and Balakrishnan, N. and Read, C. and Vidakovic, B. and Johnson, N. L.},
  title =      {Bayes Factors},
  booktitle =  {{Encyclopedia of Statistical Sciences}, vol. 1},
  edition = {2},
  address =    {Hoboken, NJ},
  publisher =  {Wiley},
  pages =      {378--386},
}

@article{steering1989final,
  title={{Final report on the aspirin component of the ongoing Physicians' Health Study}},
  author={Physicians' Health Study Research Group, Steering Committee of the},
  journal={New England Journal of Medicine},
  volume={321},
  number={3},
  pages={129--135},
  year={1989},
  publisher={Mass Medical Soc}
}

@article{joura20159,
  title={{A 9-valent HPV vaccine against infection and intraepithelial neoplasia in women}},
  author={Joura, Elmar A and Giuliano, Anna R and Iversen, Ole-Erik and Bouchard, Celine and Mao, Constance and Mehlsen, Jesper and Moreira Jr, Edson D and Ngan, Yuen and Petersen, Lone Kjeld and Lazcano-Ponce, Eduardo and others},
  journal={New England Journal of Medicine},
  volume={372},
  number={8},
  pages={711--723},
  year={2015},
  publisher={Mass Medical Soc}
}

@article{magee2015less,
  title={Less-tight versus tight control of hypertension in pregnancy},
  author={Magee, Laura A and Von Dadelszen, Peter and Rey, Evelyne and Ross, Susan and Asztalos, Elizabeth and Murphy, Kellie E and Menzies, Jennifer and Sanchez, Johanna and Singer, Joel and Gafni, Amiram and others},
  journal={New England Journal of Medicine},
  volume={372},
  number={5},
  pages={407--417},
  year={2015},
  publisher={Mass Medical Soc}
}

@INCOLLECTION{wagenmakers2020principle,
  AUTHOR =       {Wagenmakers, E.--J. and Lee, M. D. and Rouder, J. N. and Morey, R. D.},
  editor =       {Gruber, C. W.},
  booktitle =    {{The Theory of Statistics in Psychology -- Applications, Use and Misunderstandings}},
  title =        {The Principle of Predictive Irrelevance or Why Intervals Should not be Used for Model Comparison Featuring a Point Null Hypothesis},
  pages =        {111--129},
  PUBLISHER =    {Springer},
  YEAR =         {2020},
  address =      {Cham},
}

@article{ploderl2013suicide,
  title={Suicide risk and sexual orientation: a critical review},
  author={Pl{\"o}derl, Martin and Wagenmakers, Eric-Jan and Tremblay, Pierre and Ramsay, Richard and Kralovec, Karl and Fartacek, Clemens and Fartacek, Reinhold},
  journal={Archives of Sexual Behavior},
  volume={42},
  number={5},
  pages={715--727},
  year={2013},
  publisher={Springer}
}

@article{polack2020safety,
  title={{Safety and efficacy of the BNT162b2 mRNA Covid-19 vaccine}},
  author={Polack, Fernando P and Thomas, Stephen J and Kitchin, Nicholas and Absalon, Judith and Gurtman, Alejandra and Lockhart, Stephen and Perez, John L and P{\'e}rez Marc, Gonzalo and Moreira, Edson D and Zerbini, Cristiano and others},
  journal={New England Journal of Medicine},
  volume={383},
  number={27},
  pages={2603--2615},
  year={2020},
  publisher={Mass Medical Soc}
}

@article{gunel1974bayes,
  title={Bayes factors for independence in contingency tables},
  author={Gunel, Erdogan and Dickey, James},
  journal={Biometrika},
  volume={61},
  number={3},
  pages={545--557},
  year={1974},
  publisher={Oxford University Press}
}

@article{jeffreys1935some,
  title={Some tests of significance, treated by the theory of probability},
  author={Jeffreys, Harold},
  volume={31},
  number={2},
  pages={203--222},
  year={1935},
  journal={Proceedings of the Cambridge Philosophical Society}
}

@article{kass1995bayes,
  title={Bayes factors},
  author={Kass, Robert E and Raftery, Adrian E},
  journal={Journal of the American Statistical Association},
  volume={90},
  number={430},
  pages={773--795},
  year={1995},
  publisher={Taylor \& Francis}
}

@article{ly2016harold,
  title={{Harold Jeffreys’s default Bayes factor hypothesis tests: Explanation, extension, and application in psychology}},
  author={Ly, Alexander and Verhagen, Josine and Wagenmakers, Eric-Jan},
  journal={Journal of Mathematical Psychology},
  volume={72},
  pages={19--32},
  year={2016},
  publisher={Elsevier}
}

@book{jeffreys1939theory,
  title={{Theory of Probability}},
  author={Jeffreys, Harold},
  edition={1},
  year={1939},
  publisher={Oxford, UK: Oxford University Press}
}

@article{morey2016philosophy,
  title={{The philosophy of Bayes factors and the quantification of statistical evidence}},
  author={Morey, Richard D and Romeijn, Jan-Willem and Rouder, Jeffrey N},
  journal={Journal of Mathematical Psychology},
  volume={72},
  pages={6--18},
  year={2016},
  publisher={Elsevier}
}
\newpage

\appendix
\section{Induced Priors} \label{sec:induced-priors}
Under the LT approach, the marginal priors for $\theta_1$ and $\theta_2$ as well as $\eta$ are given in integral representation by \textcite{gronau2019informed}. Under the IB approach, the density of $\eta$ is given by:
\begin{equation*}
    f(\eta; a) = \begin{cases} 
    \frac{1}{B(a, a)}\eta^{2a - 1} (1 - \eta)^{2a - 1} \FAppel{a}{4a - 2}{1 - a}{2a}{1 - \eta}{1 - \eta^2} & 0 < \eta \leq 1 \\[0.5em]
    \frac{1}{B(a, a)}(-\eta)^{2a - 1} (1 + \eta)^{2a - 1} \FAppel{a}{1 - a}{4a - 2}{2a}{1 - \eta^2}{1 + \eta} & -1 \leq \eta < 0
    \end{cases}
    \enspace ,
\end{equation*}
see \textcite{pham1993bayesian}. In our case, we always have that $2a > 1$ and so $f(0) = \frac{B(2a - 1, 2a - 1)}{B(a, a)^2}$ \parencite{pham1993bayesian}. We can derive the density for $\psi$ under prior independence for the special case of $a = 1$. Note that, because $\theta \sim \text{Beta}(1, 1)$, it follows that:

\begin{equation*}
    \text{log}\left(\frac{\theta}{1 - \theta}\right) \sim \text{Logistic}(0, 1) \enspace .
\end{equation*}

\noindent Recall that:

\begin{equation*}
    \psi = \text{log}\left(\frac{\theta_2}{1 - \theta_2}\right) - \text{log}\left(\frac{\theta_1}{1 - \theta_1}\right) \enspace .
\end{equation*}

\indent \textcite{ojo2003remark} gives the distribution function of the sum of $n$ i.i.d. logistic random variables. Since the logistic distribution is symmetric, we have that if $X \sim \text{Logistic}$ then $-X \sim \text{Logistic}$. Hence we can use the results by \textcite{ojo2003remark} and write the density function of $\psi$ as:

\begin{equation*}
    f(\psi; a = 1) = \frac{e^{\psi}(e^{\psi} (\psi - 2) + \psi + 2)}{(e^{\psi} - 1)^3} \enspace .
\end{equation*}

\section{Combining Approaches through Model-averaging} \label{sec:model-averaging}
If one is uncertain as to which test one should employ, an approach would be to combine both through model-averaging \parencite[e.g.,][]{hoeting1999bayesian, hinne2020conceptual}. Let $\mathcal{M}_0^{\text{IB}}$ and $\mathcal{M}_0^{\text{LT}}$ denote the null models under the IB and the LT approach, respectively, and let $\mathcal{M}_1^{\text{IB}}$ and $\mathcal{M}_1^{\text{LT}}$ denote the respective alternative models. The model-averaged Bayes factor is given by:
\begin{equation*}
    \text{BF}^{\text{AVG}}_{01} = \frac{p(\mathcal{D} \mid \mathcal{M}_0^{\text{IB}}) \pi(\mathcal{M}_0^{\text{IB}}) + p(\mathcal{D} \mid \mathcal{M}_0^{\text{LT}}) \pi(\mathcal{M}_0^{\text{LT}})}{p(\mathcal{D} \mid \mathcal{M}_1^{\text{IB}}) \pi(\mathcal{M}_1^{\text{IB}}) + p(\mathcal{D} \mid \mathcal{M}_1^{\text{LT}}) \pi(\mathcal{M}_1^{\text{LT}})} \enspace ,
\end{equation*}
where $\pi()$ gives the prior probability of the particular model. One has to be careful when model-averaging. While the nuisance parameters generally do not matter when testing nested models \parencite{kass1992approximate, ly2016harold}, they do here. In particular, the IB approach assigns $\beta$ a logistic distribution, while the LT approach assigns $\beta$ a Gaussian distribution. The logistic distribution has fatter tails, and thus puts more mass on extreme values of $\theta$ than the Gaussian distribution. In our case, this can result in somewhat surprising results. Let $\mathcal{D}_1 = (y_1, n_1)$ and $\mathcal{D}_2 = (y_2, n_2)$ denote data in the two groups, respectively. For data $\mathcal{D}_1 = (0, 50)$ and $\mathcal{D}_2 = (0, 50)$, for example, the IB model that assumes a difference in the population ($\mathcal{M}^{\text{IB}}_1$) outpredicts the LT null model ($\mathcal{M}^{\text{LT}}_0$) by a factor of seven! Since we want to focus on differences under $\mathcal{H}_1$ between the IB and LT approach, we would need to make the models identical under $\mathcal{H}_0$, for example by assigning $\beta$ a logistic prior not only under the IB approach, but also under the LT approach.

\section{Inference based on Posterior Distributions} \label{sec:posterior-inference}
\begin{figure}[ht]
   \centering
   \includegraphics[width = 1\textwidth]{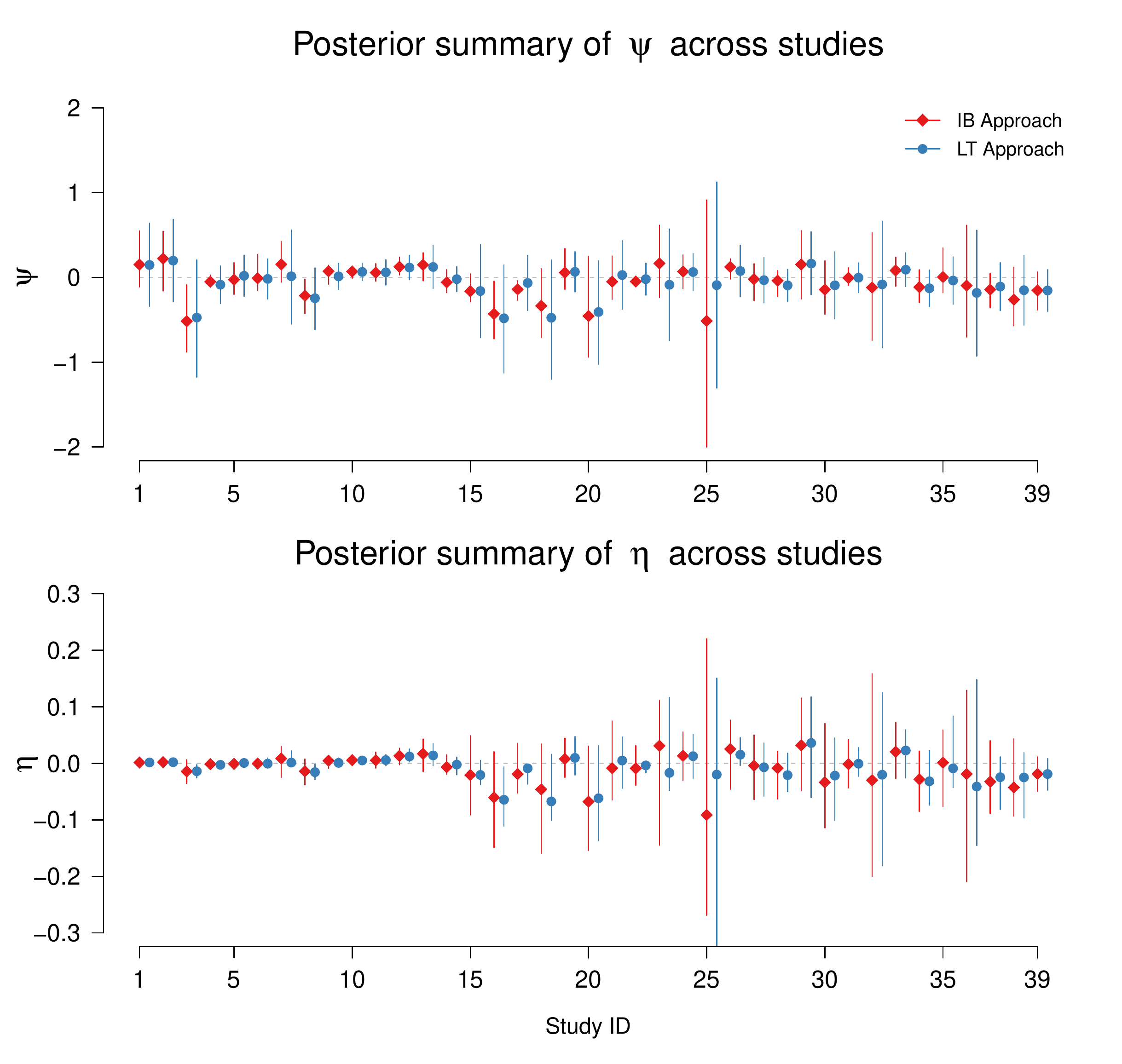}
   \caption{Shows the posterior mean and 95\% credible interval of the log odds difference $\psi$ (top) and the rate difference $\eta$ (bottom) under the IB approach (rectangles, $a = 1$) and the LT approach (circles, $\sigma_{\psi} = 1$).}
   \label{fig:ci-reanalysis}
\end{figure}

The contrasting results discussed in Section \ref{sec:paradox} do not occur when inferences are based on the posterior distribution of the log odds difference $\psi$ or the rate difference $\eta$. Figure \ref{fig:ci-reanalysis} shows the posterior mean and 95\% credible interval of $\psi$ (top) and $\eta$ (bottom) for the IB (rectangles) and the LT approach (circles) with the respective default parameterizations $a = 1$ and $\sigma_{\psi} = 1$ across all studies. The conclusions one would draw from these posterior distributions are very similar (and changing the prior parameters does not affect them). This is as expected, since the effect of the prior generally washes out as more and more data are observed, which is not the case for Bayes factors.

\section{A Dependent IB Approach} \label{sec:dependent-CT}
\begin{figure}[ht]
   \centering
   \includegraphics[width = 1\textwidth]{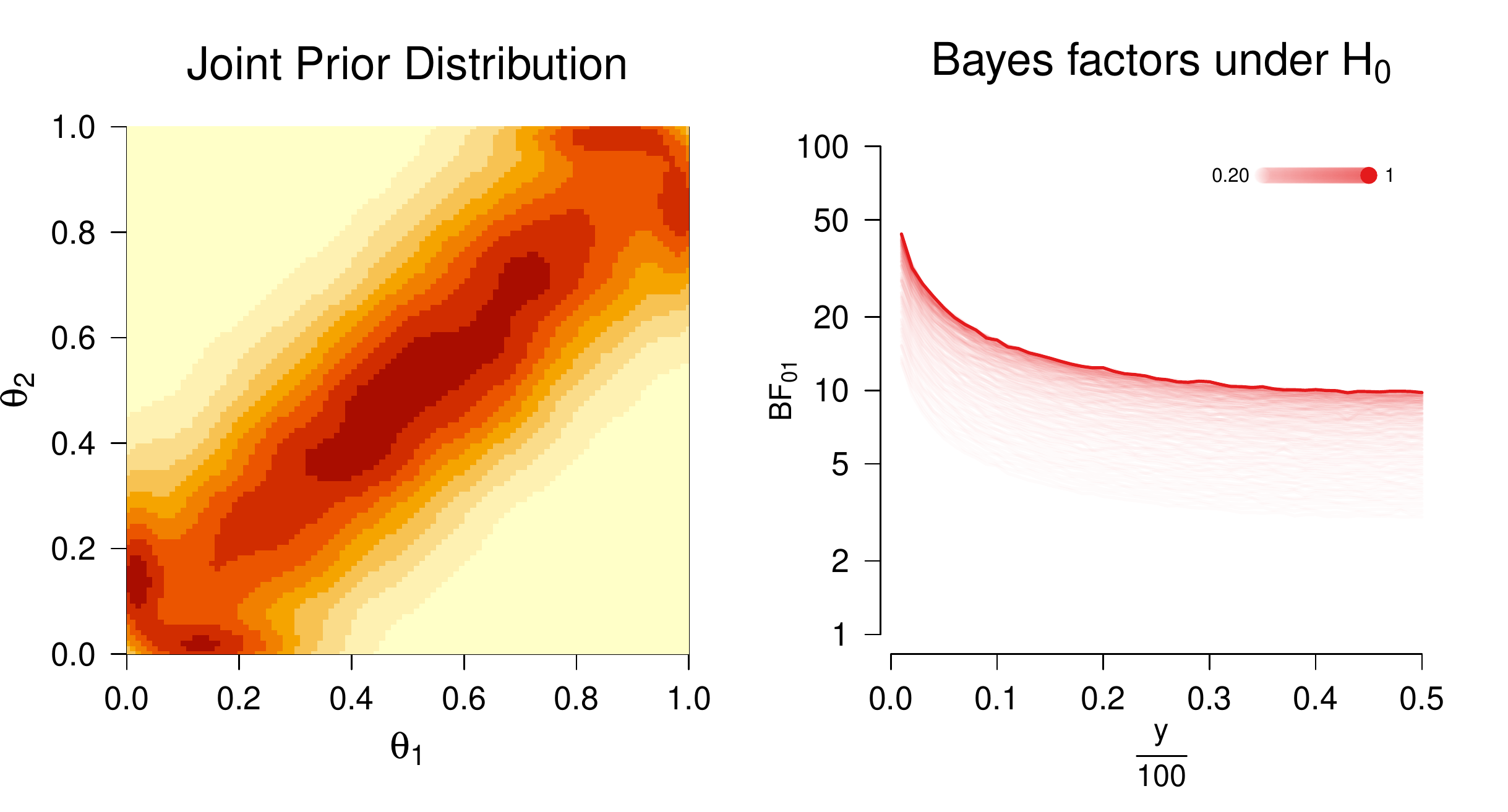}
   \caption{Left: Joint distribution on $(\theta_1, \theta_2)$ using $\sigma_{\eta} = \nicefrac{1}{5}$ and $\sigma_{\zeta} = \nicefrac{1}{2}$. Right: Bayes factors in favour of $\mathcal{H}_0$ for simulated proportions with $n = 100$ for values $\sigma_{\eta} = [\nicefrac{1}{5}, 1]$ and $\sigma_{\zeta} = \nicefrac{1}{2}$.}
   \label{fig:dependent-ct}
\end{figure}

If prior dependence were responsible for the stark differences between the Bayes factors at the extremes, we would find that a IB approach which assumes prior dependence shows a pattern similar to the LT approach. Here, we show that this is not the case. Recall that $\eta = \theta_2 - \theta_1$ and $\zeta = \frac{1}{2} \left(\theta_1 + \theta_2\right)$. We define:
\begin{align*}
  \theta_1 &= \text{min}\left(\text{max}\left(\zeta - \frac{\eta}{2}, 0\right), 1\right) \\
  \theta_2 &= \text{min}\left(\text{max}\left(\zeta + \frac{\eta}{2}, 0\right), 1\right) \enspace ,
\end{align*}
and assign truncated Gaussian priors to $(\eta, \zeta$), i.e., $\eta \sim \mathcal{N}(0, \sigma_{\eta})_{\mathbb{I}(-1, 1)}$ and $\zeta \sim \mathcal{N}(0, \sigma_{\zeta})_{\mathbb{I}(0, 1)}$. Using $\sigma_{\eta} = \nicefrac{1}{5}$ and $\sigma_{\zeta} = \nicefrac{1}{2}$, the left panel in Figure \ref{fig:dependent-ct} shows the joint prior distribution for $(\theta_1, \theta_2)$ from which a strong prior dependence between $\theta_1$ and $\theta_2$ is apparent. The right panel, however, shows the characteristic pattern of the IB approach as discussed in the main text, rather than the characteristic pattern of the LT approach. Since the rates a dependent a priori, it is not the prior dependence that is responsible for the pattern. However, employing a prior that has a stronger dependency between the two rates naturally reduces the size of their expected difference. Increasing $\sigma_{\eta}$ results in a lower correlation (going from about 0.77 at $\sigma_{\eta} = \nicefrac{1}{5}$ to zero at $\sigma_{\eta} = 1$) and a reduced Bayes factor in favour of $\mathcal{H}_0$, as the right panel in Figure \ref{fig:dependent-ct} shows.

\section{Sequential Predictive Perspective} \label{sec:prequential}
Under $\mathcal{H}_0$ we have for both the IB (i.e., independent) and the LT (i.e., dependent) setup that (suppressing conditioning on $\mathcal{H}_0$):
\begin{align}
    p(\mathcal{D}_2, \mathcal{D}_1 \mid \mathcal{H}_0) &= \int_{\theta} p(\mathcal{D}_2, \mathcal{D}_1 \mid \theta) \pi_0(\theta) \, \, \mathrm{d}\theta \\[0.50em]
    &= \int_{\theta} p(\mathcal{D}_2 \mid \mathcal{D}_1, \theta) p(\mathcal{D}_1 \mid \theta) \pi_0(\theta) \, \, \mathrm{d}\theta \\[0.50em]
    &= \mathcal{Z} \int_{\theta} p(\mathcal{D}_2 \mid \theta) p(\theta \mid \mathcal{D}_1) \, \, \mathrm{d}\theta \enspace ,
\end{align}
where $\mathcal{Z} = \int_{\theta} p(\mathcal{D}_1 \mid \theta) \pi_0(\theta) \, \, \mathrm{d}\theta$ is the marginal likelihood for $\mathcal{D}_1$, $p(\theta \mid \mathcal{D}_1)$ is the posterior of $\theta$ after observing $\mathcal{D}_1$, and we can remove the conditioning on $\mathcal{D}_1$ in (2) because all relevant information is in $\theta$. Similarly, under $\mathcal{H}_1$ for LT setup we have that:
\begin{align}
    p(\mathcal{D}_2, \mathcal{D}_1 \mid \mathcal{H}_1) &= \int_{\theta_2} \int_{\theta_1}p(\mathcal{D}_2, \mathcal{D}_1 \mid \theta_1, \theta_2) \pi_1(\theta_1, \theta_2) \, \mathrm{d}\theta_1\, \mathrm{d}\theta_2 \\[0.50em]
    &= \int_{\theta_2} \int_{\theta_1} p(\mathcal{D}_2 \mid \mathcal{D}_1, \theta_1, \theta_2) p(\mathcal{D}_1 \mid \theta_1, \theta_2) \pi_1(\theta_1, \theta_2) \, \mathrm{d}\theta_1 \, \mathrm{d}\theta_2 \\[0.50em]
    &= \mathcal{Z} \int_{\theta_2} \int_{\theta_1} p(\mathcal{D}_2 \mid \theta_1, \theta_2) p(\theta_1, \theta_2 \mid \mathcal{D}_1) \, \mathrm{d}\theta_1 \, \mathrm{d}\theta_2 \enspace ,
\end{align}
where $\mathcal{Z}$ is the marginal likelihood for $\mathcal{D}_1$, $p(\theta_1, \theta_2 \mid \mathcal{D}_1)$ is the posterior after observing $\mathcal{D}_1$, and we can remove the conditioning on $\mathcal{D}_1$ in (2) because all relevant information is in $(\theta_1, \theta_2)$. This is in contrast to the IB setup, where:
\begin{align}
    p(\mathcal{D}_2, \mathcal{D}_1 \mid \mathcal{H}_1) &= \int_{\theta_2} \int_{\theta_1}p(\mathcal{D}_2, \mathcal{D}_1 \mid \theta_1, \theta_2) \pi_1(\theta_1, \theta_2) \, \mathrm{d}\theta_1\, \mathrm{d}\theta_2 \\[0.50em]
    &= \int_{\theta_2} p(\mathcal{D}_2 \mid \theta_2) \pi_1(\theta_2) \, \mathrm{d}\theta_2 \int_{\theta_1}p(\mathcal{D}_2 \mid \theta_1) \pi_1(\theta_1) \, \mathrm{d}\theta_1 \enspace ,
\end{align}
and thus no sharing of information across the two groups takes place.
\end{document}